\RequirePackage{fix-cm}
\documentclass[final,3p,twocolumn,authoryear]{elsarticle}
\usepackage{graphicx}
\usepackage{mathptmx}      % use Times fonts if available on your TeX system
\usepackage{amsmath}
\usepackage{natbib}
\usepackage{xcolor}
\usepackage[unicode,pdfhighlight=/P,colorlinks=true]{hyperref}
\newcommand{\Geo}[2]{\ensuremath{#1^\circ\,\text{#2}}}
\newcommand{\E}[1]{\Geo{#1}{E}}
\newcommand{\W}[1]{\Geo{#1}{W}}
\newcommand{\N}[1]{\Geo{#1}{N}}

\begin{document}
\hypersetup{colorlinks=true,linkcolor=red,citecolor=brown,urlcolor=blue,unicode,pdfhighlight=/P}
\begin{frontmatter}
\title{Lagrangian study of temporal changes of a surface flow through the Kamchatka Strait}


\author{S.V. Prants}
\ead{prants@poi.dvo.ru}
\author{A.G. Andreev}
\author{M.Yu. Uleysky}
\author{M.V. Budyansky}


\address{Pacific Oceanological Institute of the Russian Academy of Sciences,\\
43 Baltiiskaya st., 690041 Vladivostok, Russia\\
URL: \url{http://dynalab.poi.dvo.ru}}

\begin{abstract}
Using Lagrangian methods we analyze a 20-year-long estimate of water flux through
the Kamchatka Strait in the northern North Pacific based on AVISO velocity field.
It sheds new light on the flux pattern and its
variability on annual and monthly time scales. Strong seasonality in 
surface outflow through the strait could be explained by temporal changes
in the wind stress over the northern and western Bering Sea slopes. Interannual
changes in a surface outflow through the Kamchatka Strait
correlate significantly with the Near Strait inflow and Bering Strait
outflow.
Enhanced westward surface flow of the Alaskan Stream across the \E{174}
section
in the northern North Pacific is accompanied by an increased inflow into the
Bering Sea through the Near Strait. In summer, the surface flow pattern in
the Kamchatka Strait is determined by passage of anticyclonic and cyclonic
mesoscale eddies. The wind stress over the Bering basin in winter~-- spring is 
responsible for eddy generation in the region.
\end{abstract}
\end{frontmatter}

\section{Introduction}\label{intro}
\begin{figure}
\begin{center}
\includegraphics[width=0.49\textwidth,clip]{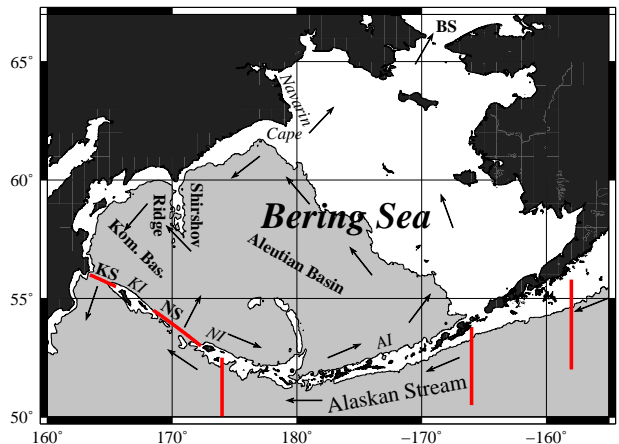}
\end{center}
\caption{Schematic representation of the currents in the study area. 
Kom. Bas.~--- Komandorsky Basin, BS~--- Bering Strait,
KS~---
Kamchatka Strait, NS~--- Near Strait, AI~--- Aleutian Islands, KI~--- Komandorsky Islands,
NI~--- Near
Islands. Thick straight lines show the sections where surface flow
anomalies were estimated. Depths more than 1000~m are shaded by grey.}
\label{fig1}
\end{figure}

The water circulation in the Bering Sea is tightly connected with a general
circulation in the northern North Pacific. It is formed by the cyclonic gyre
with two main currents, the  Kamchatka Current along the western boundary and the Bering
Slope Current in the eastern part of the sea. The most part of Pacific water enters 
the Bering Sea through the Amukta Pass \citep{Stabeno05,Stabeno09} and Near Strait 
(sill depth is about 2000~m) (Fig.~\ref{fig1}). The Alaskan Stream, transporting relatively warm
subsurface waters, is the main source of waters in the upper 1000~m layer in the
Bering Sea. The water outflow from the Bering Sea into the Pacific Ocean occurs mainly through
the Kamchatka Strait.
The Kamchatka Strait (KS) with its width of 190~km and maximum depth of 4420~m, located between the
Kamchatka Peninsula and the Bering Island,
occupies 46\% of the total area of the straits connecting the Bering Sea and the Pacific Ocean.
The water transport through the KS
and straits of the Aleutian Island chain is of great importance for the volume, heat and
nutrient fluxes between the Bering Sea
and the North Pacific \citep{Stabeno99}. According to various estimates
\citep{Overland94,Reed95,Cokelet96,Panteleev06}, the volume transport makes up from 5
to 25~Sv ($\text{1~Sv}=10^6$~m$^3$/s). Speeds of geostrophic flows in the strait, 25--40~cm s$^{-1}$, have the maximum in the upper layer
\citep{Cokelet96}. Bering Strait (55~m deep and 85~km wide) provides a connection between the
Bering Sea and the Chukchi Sea  in the Arctic Ocean.
The northward flow through the Bering Strait (${\sim}1$~Sv) is insignificant to the water
budget of the deep Bering Sea but it helps balance the
freshwater budgets of the Pacific and Atlantic oceans.

Mesoscale variability is an important factor in the Bering Sea dynamics. 
The eddy activity in the vicinity of Near Strait has a strong impact on 
the flow through the strait \citep{Stabeno93}. Generation of anticyclonic and cyclonic submesoscale eddies in Near Strait leads to the flow reversals in central and eastern 
parts of the strait and westward (or eastward) shift in the velocity core \citep{OM13}.
The meanders and eddies are consistent features in the Bering
Slope Current and the Kamchatka Current regions \citep{Verkhunov92,Mizobata02}. The
distributions of the sea surface height anomaly (SSHA) in
the KS area and satellite infrared images indicate strong anticyclonic and cyclonic eddy
activity in the strait.
Although the flux pattern in the KS has been known for decades, volume transport variability
and its causes remain highly
uncertain because of the lack of direct observations. In this study we analyze a 20-year-long
estimate of the water flux through the KS based on AVISO surface velocity fields.

To estimate surface flow in the region we apply in this paper the Lagrangian approach that is based
on computing trajectories of passive particles
advected by an AVISO velocity field. The domain under study is seeded by a large number of synthetic
particles whose trajectories are computed forward or backward in time for a given period of time.
When integrating advection equations (\ref{adveq}) forward
in time one gets an information about fate of water masses and when integrating them backward in time
we can determine their origin and history.
A graphic view of transport and mixing in a studied area is provided by so-called
Lagrangian maps which are plots of one of the Lagrangian indicators versus particle's initial
positions \citep{Chaos17,DAN11,OM11,P13,FAO13}. That approach has been successfully applied to
study transport and mixing in different basins,
from marine bays \citep{FAO13} and seas \citep{OM11,OM13} to the ocean scale \citep{DAN11,P13}.

The Lagrangian approach sheds new insight on the flow patterns in the KS and the variability 
on annual to monthly time scales. 
Computed backward-in-time latitudinal Lagrangian maps
show mesoscale and submesoscale eddies in the region and origin and
history of water masses crossing the strait for a given period of time.
We demonstrate in this paper correlations between the outflow flux through the KS,
the inflow through the Near Strait, the Bering Strait transport derived from moored
observations \citep{Woodgate12} and Alaskan Stream surface flow.

\section{Data and methods}\label{data_methods}
Since 1992, the ocean surface topography has been continuously observed from the space by the
Topex/Poseidon, European Remote Sensing,
Geosat Follow-On, Envisat and Jason satellite missions. These data are available at
\url{http://www.aviso.oceanobs.com}. We use the
gridded SSHA for the period of 1993--2012 for diagnostic computations. For each 7-day
period, the SSHAs, obtained by optimal interpolation
of all available altimeter missions, were downloaded from the AVISO site.   In our study,
we use the monthly wind stress dataset from the
NCEP reanalysis. The horizontal resolution of the NCEP data is $1.9^\circ\times1.9^\circ$.

Geostrophic daily velocities, obtained from the AVISO database on
a $1/3^\circ\times1/3^\circ$ Mercator grid, are an approximation to the real surface fields
in the ocean. To provide accurate numerical results we apply a bicubical spatial interpolation
and third order Lagrangian polynomials in time.
We compute in this paper transport for periods as long as a few weeks. 
It is reasonably to suppose that various ageostrophic corrections
are averaged out for such a long time. Moreover, our results are based not on
individual trajectories but on statistics
for thousands of trajectories. Most of them have been found to be
chaotic with positive values of the maximal finite-time Lyapunov exponent
which is an average rate of dispersion of initially closed particles
\citep[see, for example,][]{OM11}.
Typical chaotic systems are highly robust against a high-frequency noise
mimicking small-scale diffusion \citep[see, e.~g.,][]{PRE06}. Individual trajectories are sensitive to
small noisy variations in the velocity field but different statistical characteristics and
structures like mesoscale manifolds and fronts are not \citep{Hernandez11,Keating11}.
Thus, an additional small noise in the advection equations is not expected
to change our results based on statistics for a large number of particles.

Lagrangian trajectories have been computed by integrating the advection equations with a
fourth-order Runge-Kutta scheme
\begin{equation}
\dot x=u(x,y,t),\qquad \dot y=v(x,y,t),
\label{adveq}
\end{equation}
where time $t$ is in days. Coordinates $x$ and $y$ of a passive particle
are related with its latitude $\phi$ and longitude $\lambda$
in degrees as follows:
\begin{equation}
\begin{gathered}
\lambda=\frac{x}{60},\quad
\phi=\frac{180}{\pi}\arcsin{\tanh{\left(\frac{\pi}{180}\left(\frac{y}{60}
+y_0\right)\right)}},\\
 y_0=\frac{180}{\pi}\operatorname{artanh}{\sin{\left(\frac{\pi}{180}\phi_0\right)}},\quad
\phi_0=-82.
\end{gathered}
\label{coordinate}
\end{equation}
We use the transformation (\ref{coordinate}) because the AVISO grid is homogeneous in
$x\text{--}y$ coordinates. The velocities $u$ and $v$ in Eq.~(\ref{adveq}) are
expressed through
the latitudinal $U_\phi$ and longitudinal $U_\lambda$ components of the linear velocity
$U$ in cm s$^{-1}$ as follows:
\begin{equation}
\begin{aligned}
u&=\frac{10800}{\pi R_E\cos\phi}\frac{86400}{100000}U_\lambda\approx
\frac{0.466}{\cos\phi}U_\lambda,\\
v&=\frac{10800}{\pi R_E\cos\phi}\frac{86400}{100000}U_\phi\approx \frac{0.466}{\cos\phi}U_\phi,
\end{aligned}
\label{velocity}
\end{equation}
where $R_E=6370$~km is the Earth radius.

The surface fluxes through the
KS and the Near Strait have been computed as follows. The lines, crossing the straits from
$[\E{163.4}:\N{56.0}]$ to $[\E{165.5}:\N{55.5}]$
and from $[\E{168.5}:\N{54.5}]$ to $[\E{172.3}:\N{53.0}]$, are divided in 20 segments. The flux across
each segment was obtained by integrating velocity normal to it at a given point. 
%The quantity we get is a daily surface flux across each segment in m$^2$/s.

\section{Results}\label{results}
\begin{figure*}[!htb]
\begin{center}
\includegraphics[width=0.8\textwidth,clip]{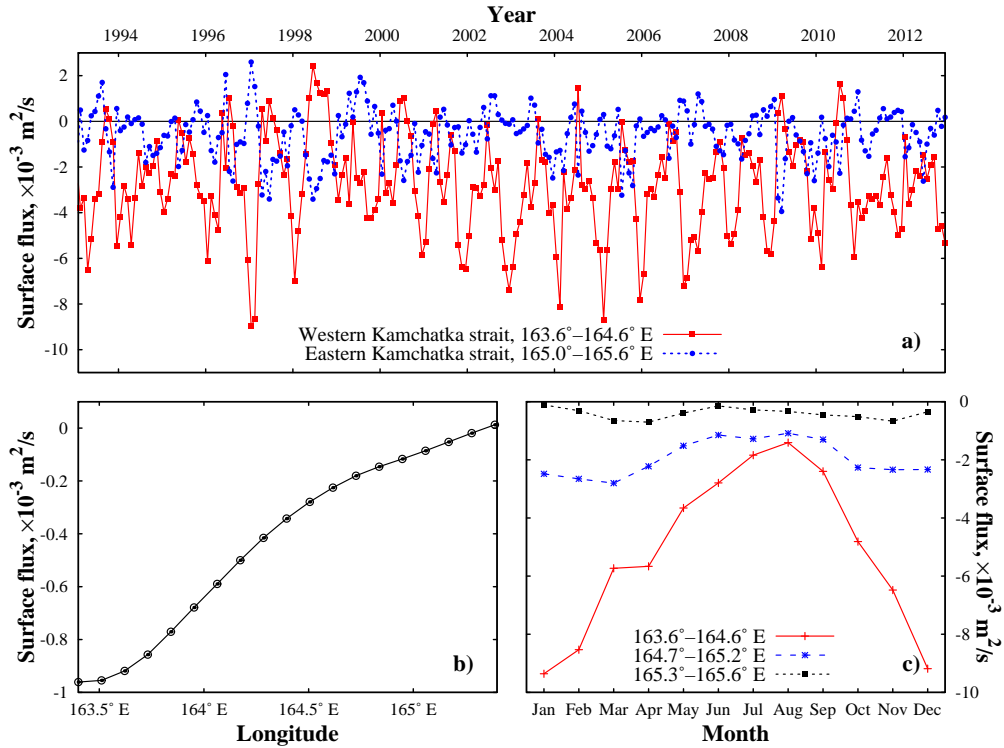}
\end{center}
\caption{a)~Temporal changes in the Kamchatka Strait fluxes, b)~flux averaged for
1993--2012 yrs vs longitude, where middles of the intervals are denoted by points, c)~monthly averaged flux through different parts of the
Kamchatka Strait.}
\label{fig2}
\end{figure*}
\begin{figure*}[!htb]
\begin{center}
\includegraphics[width=0.8\textwidth,clip]{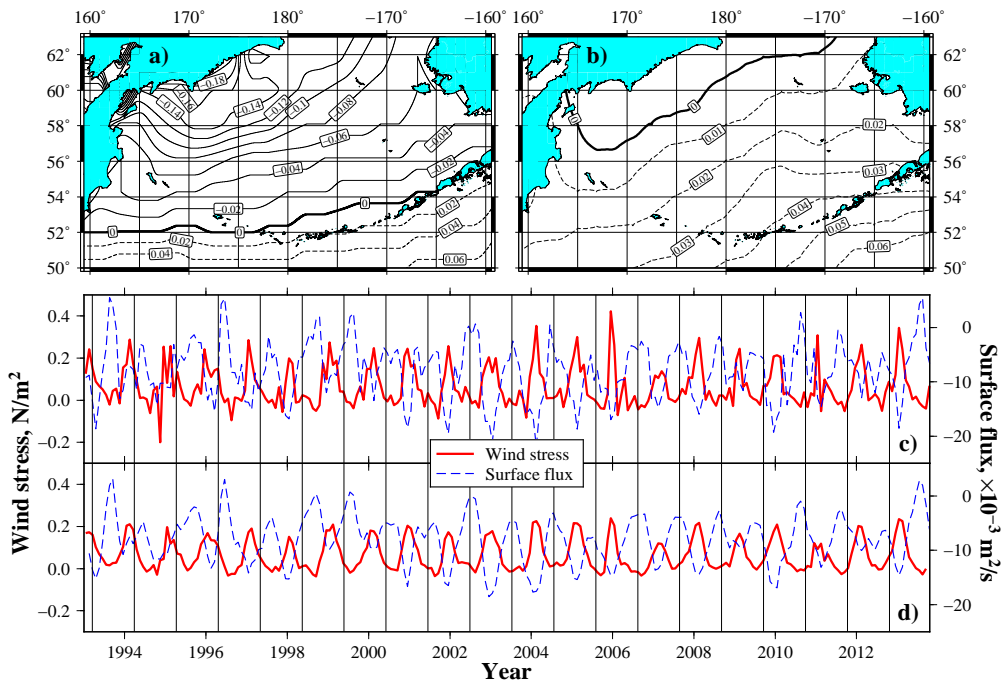}
\end{center}
\caption{Spatial patterns of zonal wind stress (N m$^{-2}$) in the northern North Pacific and
Bering Sea in a) November~-- March and b)~July~-- September 
c)~and d)~Time series of the surface flux through the Kamchatka
Strait and the wind stress over the Bering Sea
($\tau_x^{\N{58},\,\E{165}\text{--}\W{170}}+0.2\tau_y^{\N{56\text{--}58},\,\E{165}}$).
Low-pass (three-months running mean) filtered series are shown in d).}
\label{fig3}
\end{figure*}
\begin{figure*}[!htb]
\begin{center}
\includegraphics[width=0.8\textwidth,clip]{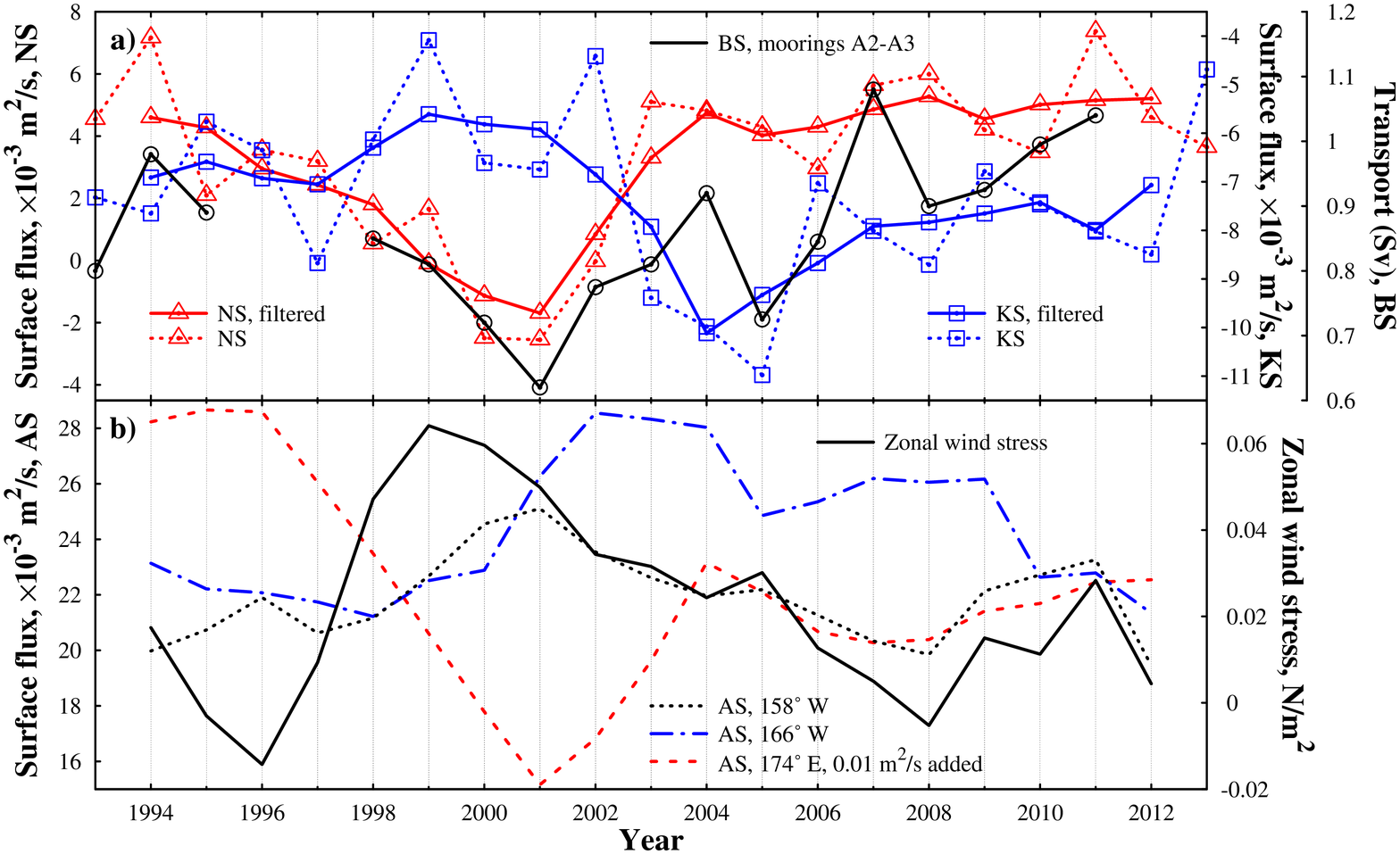}
\end{center}
\caption{a)~The interannual changes of the flux through the Kamchatka and Near Straits and the
volume transport through the Bering Sea
(\protect\url{http://psc.apl.washington.edu/HLD/Bstrait/Data}
b)~the interannual changes of the surface flux (westward is positive)
of the Alaskan Stream (AS) and the zonal wind stress over the northern
North Pacific (\N{52}, \E{165}~-- \W{170}) averaged
for November~-- March}.
\label{fig4}
\end{figure*}
Figures~\ref{fig2}a--c show monthly time series of surface fluxes through the western and
eastern parts of the KS,
the water flux averaged for 1993--2012 years versus longitude and the monthly averaged flux
through the different parts
of the KS. The computed net flow through the KS is directed from the Bering Sea to the North
Pacific. 
The northward (southward) flow is positive (negative). The surface fluxes of the Alaskan Stream were computed as flows across
\E{174} (from \N{50.0} to \N{52.5}), \W{166}
(from \N{51.5} to \N{53.5}) and \W{158} (from \N{53.0} to \N{55.0}) lines (Fig.~\ref{fig1}).
The surface southward
flow is enhanced in the western part of the strait, but the flow in its eastern side is
relatively weak (Fig.~\ref{fig2}b).
The flux time series through the western and eastern KS are negatively correlated at the 95\%
significance level (Fig.~\ref{fig2}a).
The correlation coefficient is $-0.52$ for the monthly mean values. The negative correlation
means that when the southward flow via
the western KS is strong, the flow via the eastern KS tends to be northward. A remarkable
feature of the surface flux through
the KS is relatively high amplitude of its seasonal variations (Fig.~\ref{fig2}c). The
southward flux through the western part
of the KS is strong between November and April and relatively weak in June~-- September.
\begin{figure*}[!htb]
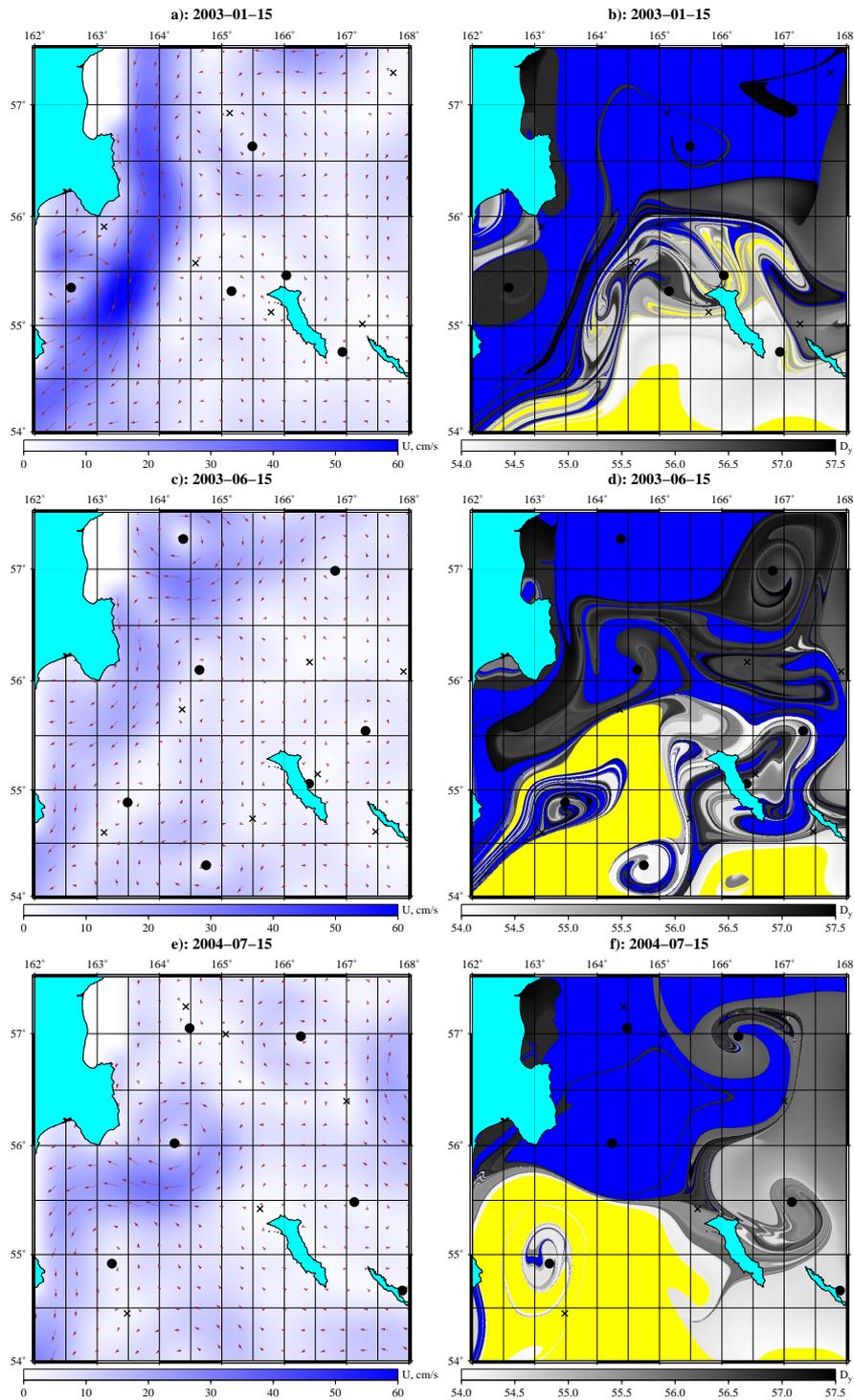

\begin{center}
\includegraphics[width=0.36\textwidth,clip]{fig5a.eps}
\includegraphics[width=0.36\textwidth,clip]{fig5b.eps}\\
\includegraphics[width=0.36\textwidth,clip]{fig5c.eps}
\includegraphics[width=0.36\textwidth,clip]{fig5d.eps}\\
\includegraphics[width=0.36\textwidth,clip]{fig5e.eps}
\includegraphics[width=0.36\textwidth,clip]{fig5f.eps}\\
\end{center}
\caption{Altimetric velocity fields (a,~c and~e) and the corresponding latitudinal Lagrangian maps
(b,~d and~f) in the Kamchatka Strait area in winter and summer. Nuances of color on velocity maps
measure the modulus of the linear velocity $U$ in cm s$^{-1}$. Hyperbolic and elliptic
stagnation points are marked by crosses and circles, respectively.
The latitudinal maps are computed backward-in-time starting from the date shown to the date
which is 180 days before.
Dark grey and black colors in geographic degrees code the
particles originated from the latitudes north of the strait, ${>}\N{58}$,
whereas the light grey and white ones code the particles coming from the
latitudes south of the strait, ${<}\N{54}$.
(In the web version of this article an additional information is coded
by ``yellow''and ``blue''
colors which mark waters that originated outside the domain and advected
from the south and north, respectively).}
\label{fig5}
\end{figure*}

The effect of local wind on the annual variation of the surface flux through the KS may be
important due to strong seasonality of
the wind forcing (Figs.~\ref{fig3}a,b). There is a tendency to have negative (westward and
southward) wind stresses over the central
and western Bering Sea in winter (Fig.~\ref{fig3}a). This wind pattern is favorable for a
pile up of the Ekman transport to set up the
alongshore jet \citep{Csanady78,Csanady98} and a sea level rise along the northern and western
boundaries of the Bering Sea basin
(the Aleutian Basin and the Komandorsky Basin). In summer, the zero wind stress contour extends
from the west-southwest to the east-northeast
across the central Bering Sea with negative values to its north and positive values to its
south (Fig.~\ref{fig3}b). Seasonal Ekman forcing is one
of the possible causes for a winter maximum of the surface flux through the KS. To assess the
effect of the Ekman forcing,
we plot in Figs.~\ref{fig3}c and 3d the monthly time series of the surface flux through the KS and the
wind stress averaged along
the northern (from \E{165} to \E{170} along \N{58}) and western (from \N{56} to \N{58} along
\E{165}) Bering Sea slope segments
($\tau_x^{\N{58},\,\E{165}\text{--}\W{170}}+0.2\tau_y^{\N{56\text{--}58},\,\E{165}}$, where
$0.2$ is the ratio between the length
of the western and northern Bering Sea slope segments). The similar approach was used for an
examination of the SSHA along the Alaska/Canada
coast from the TOPEX/Poseidon data \citep{Qiu02} and the tide gauge sea level data along the
northern and western coasts of the Okhotsk Sea
\citep{Nakanowatari13}. Our results show a good correspondence ($r=-0.59$, 
unfiltered series in Fig.~\ref{fig3}c) and $-0.69$ (low-pass filtered series
Fig.~\ref{fig3}d) between the monthly time-series
signals of the surface flux through the KS
and the wind stress $\tau_L$. The monthly time series of the volume transport along the
continental slope of the KS
(computed as $\tau_LL/(f\rho)$, where $L$ is the total length of the northern and western
Bering basin segments, $\rho$ is the density of water
and $f$ is the Coriolis parameter) exhibit a strong seasonal cycle with peak southward
(negative) transport of 6--8~Sv in winter and
near-zero or positive (${\sim}2$~Sv) transport in summer.

The bulk of inflow into the Bering Sea occurs through the Near Strait at the western end of
the Aleutian Islands (\E{168}~-- \E{172}).
Figure~\ref{fig4}a demonstrates year-to-year changes of the water flux through the KS (averaged
for January~-- September) and the Near Strait,
and the volume transport through the Bering Strait derived from 20-year moored observations
\citep{Woodgate12}. Enhanced water inflow
from the Pacific Ocean to the Bering Sea through the Near Strait accompanies increased
outflow through the KS ($r=-0.67$, 1993--2012, low-pass filtered series)
and the Bering Strait ($r=0.78$, 1993--2011). On the interannual time scale, the positive
correlation ($r=0.48$)
is diagnosed between the flow into the Bering Sea through the Near Strait and the Alaskan
Stream surface flow through the \E{174} section (Fig.~\ref{fig4}b).
Stronger Alaskan Stream in the Near Islands area appears to increase NS inflow. But there
is a negative correlation between the NS inflow
and the Alaskan Stream surface flow through the \W{158} section ($r=-0.67$).

In winter, the surface flow pattern in the KS is determined by a strong stream
directed from
the Bering Sea to the Pacific Ocean via the western part of the strait
(see the altimetric velocity field on 15 January, 2003 in Fig.~\ref{fig5}a).
In order to get a more detailed information about transport through KS,
we compute latitudinal Lagrangian maps. The domain under study is seeded with
a large number of synthetic particles which are advected backward in time
by the AVISO velocity field starting from a given day to a day in the past.
The latitude from which each particle came to its final position during that period of time
is coded by color. Figure~\ref{fig5}b shows
meridional displacements of markers during the second half of 2002
with dark grey and black colors in geographic degrees coding the
particles originated from the latitudes north of the strait (${>}\N{58}$),
whereas the light grey and white colors code the particles coming from the
latitudes south of the strait (${<}\N{54}$).
Hyperbolic and elliptic stagnation points, where the velocity is found to be
zero, are marked in this figure
by crosses and circles, respectively. Motion around ``trivial'' elliptic
points is stable, and they are situated, as a rule, in the centers of eddies.
Unstable hyperbolic or saddle points organize fluid motion in their
neighbourhood in a specific way: particles approach that point along two
stable directions and quit it along the other two unstable directions.

In summer, according to cruise data \citep{Verkhunov92}, satellite
altimetry data and infrared satellite images, the flow pattern through the KS
is induced mainly by mesoscale eddies in the strait. 
In June 2003, the cyclonic eddy perturbed the surface flow in KS, enhancing
the southward flow in the middle of the strait and causing northward flow on
the western side. The net flux through the KS was directed from the Bering
Sea into the North Pacific (Figs.~\ref{fig5}c and d). 
In summer of the next year, a mesoscale anticyclone in KS centered at
$[\E{164.25}:\N{56}]$ and
a mesoscale cyclone $[\E{163.25}:{\approx}\N{55}]$ to the south blocked 
flow through the strait (Figs.~\ref{fig5}e and f).

\begin{figure*}[!htb]
\begin{center}
\includegraphics[width=0.8\textwidth,clip]{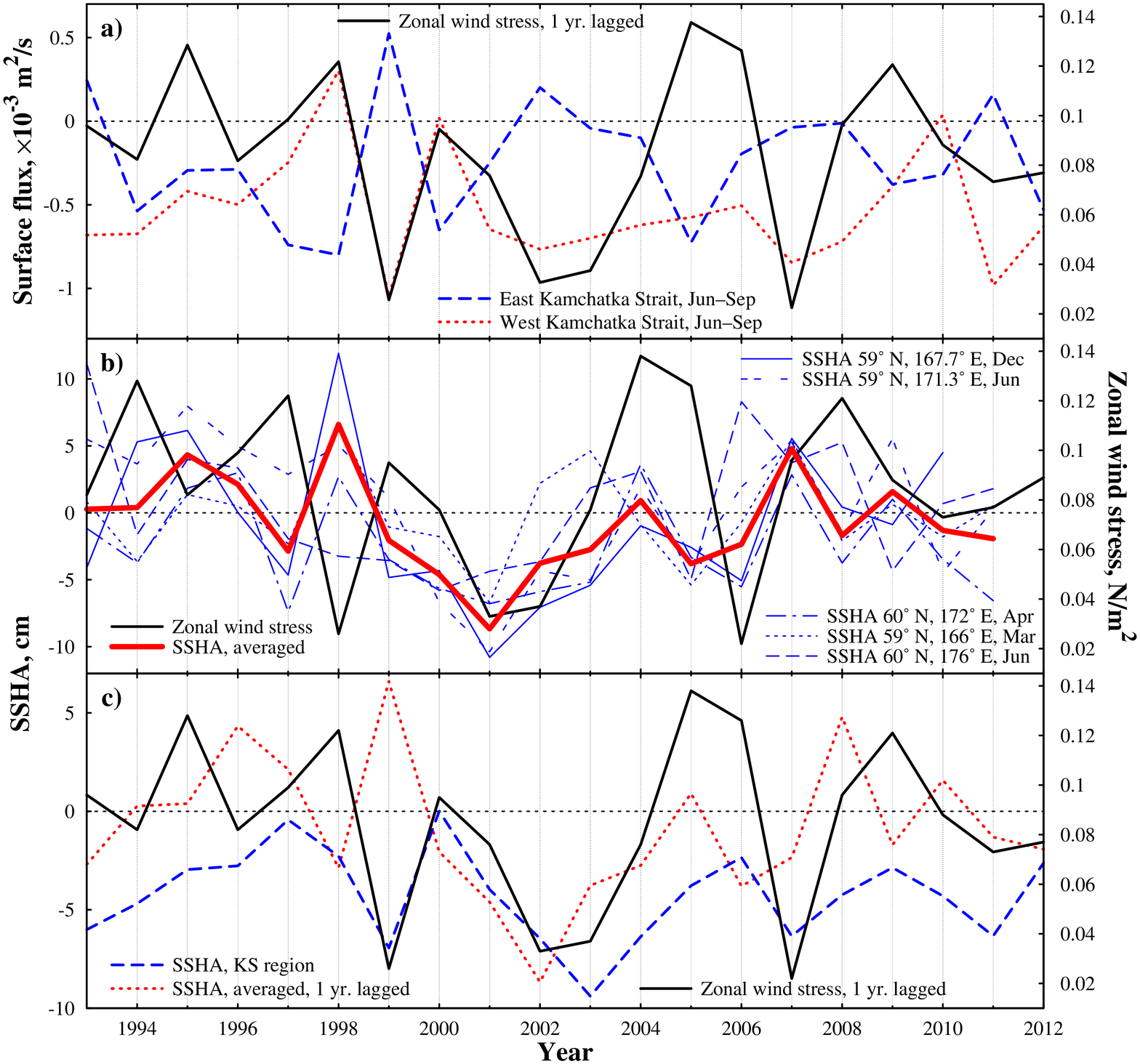}
\end{center}
\caption{Year-to-year changes of the surface flux through the Kamchatka Strait in June~-- September
(a), sea surface height anomaly (SSHA) in the Bering Sea (b and c) and the zonal wind
stress (\N{56}~-- \N{58}, \E{165}~-- \W{170}) over the Bering Sea averaged for January~-- April
(a--c).}
\label{fig6}
\end{figure*}

The flow, directed from the ocean to the Bering Sea in the eastern
part of the KS, is accompanied by a flow, directed
from the Bering Sea to the ocean in the western part of the KS ($r=-0.74$),
and {\em vice versa}
(Fig.~\ref{fig6}a). The year-to-year changes
in the fluxes through the western and eastern parts of the KS in June~-- September are correlated
($r= 0.55$ and $-0.57$) with the 1-year
lagged zonal wind stress over the Bering basin,
$\tau_x^{\N{58},\,\E{165}\text{--}\W{170}}$, in January~-- April (Fig.~\ref{fig6}a).
Increased/decreased westward component of the wind stress over the Bering basin in winter leads
to generation of negative/positive
SSHAs along the Bering slope off the Navarin Cape and in Shirshov Ridge area ($r=0.57\text{--}0.79$)
(Fig.~\ref{fig6}b). These SSHAs could
be observed in KS area with a 1-year lag (Fig.~\ref{fig6}c). The correlation coefficient
between the SSHA time-series
in KS area and the 1-year lagged SSHA along the northern and western Bering slope in spring~-- fall is
$0.84$.
\section{Discussion}\label{discuss}
\subsection{Spatial and temporal variation of the surface flow}

The annual modulation in the Bering Sea gyre intensity may be interpreted as a barotropic
response to seasonal wind stress curl forcing.
The magnitude of this response is quantifiable by a time-dependent Sverdrup balance
\citep{Bond94}.
Our results (Fig.~\ref{fig3}c) demonstrate
that seasonal wind stress forcing is one of possible reasons for the winter maximum of the
surface flux through the KS.
Based on results of a numerical model simulation, it has been found that monthly
mean volume flux time series through the
Near Strait and the KS were significantly correlated with the correlation
coefficient $-0.80$ \citep{Kinney12}. The correlation
essentially means that when inflow into the Bering Sea via the Near Strait is strong,
outflow via the KS also tends to be
strong and {\em vice versa}. Correlations between volume transport through the KS
and the passes east of the Near Strait are
much lower in magnitude (from $0.15$ to $0.27$) and, therefore, less important than
correlation between the KS and the Near Strait
\citep{Kinney12}. On the interannual scale, we diagnosed the statistically significant
correlation ($r=-0.67$, 1993--2012) between
inflow into the Bering Sea through the Near Strait and outflow through the KS
(Fig.~\ref{fig4}a). On the seasonal time scale,
we did not observe a statistically significant correlation between Near Strait inflow and
outflow via the KS. The main outflow through the KS occurs
from November to April (Fig.~\ref{fig2}c). But the monthly averaged inflows via the Near Strait
in winter are only 20 percent higher than those in summer.

The changes in the Alaskan Stream transport may be one of the main causes of variability in
water inflow via the Aleutian Passes and outflow
through the KS. Based on results of the ocean circulation model simulation, it has been
demonstrated that an increase of the
Alaskan Stream transport from 10 to 25~Sv causes warming and sea level rise in the Bering Sea
shelf due to increased transport of warmer
Pacific waters through the eastern passages of the Aleutian Islands \citep{Ezer10}. An increase
of the Alaskan Stream transport from 25 to 40~Sv had an
opposite impact on the Bering Sea  shelf with a slight cooling \citep{Ezer10}. By using SSHA
observed by satellites in 1992--2010 and monthly
climatology of temperature and salinity, it has been found by \citet{Panteleev12}
a significant negative correlation
($r=-0.84$) between the low-pass filtered series
of the Near Strait inflow and the Alaskan Stream transport across the \W{158} section.
The stronger Alaskan Stream appears to reduce the Near Strait
inflow and at the same time produces a larger transport through the Aleutian Arc which
amplifies the main cyclonic gyre controlled by the
continental slope within the Bering Slope Current region \citep{Panteleev12}.

The surface anomalies of the Near Strait inflow and the Alaskan Stream flow across \W{158}
computed by us are in a good agreement with the transport
anomalies calculated by \citet{Panteleev12}. On the interannual scale, we obtain a negative
correlation between the Alaskan Stream flow across \W{158} and the Near
Strait inflow ($r=-0.67$). However, our results demonstrate (Figs.~\ref{fig4}a and b) that
there is a positive correlation between the Alaskan
Stream flow across \E{174} (the Near Islands area) and the Near Strait inflow ($r=0.48$).
It means that the Near Strait inflow is amplified when the
Alaskan Stream in the Near Islands area is relatively strong.  There is a significant difference
between the time-series of the Alaskan Stream
flux across \E{174} (the western Aleutian Islands area), \W{166} (the eastern Aleutian Islands
area) and \W{158} (the Alaska region) (Fig.~\ref{fig4}b).
The \W{158} section crosses the northern boundary of the Alaska cyclonic gyre. The difference
in the Alaskan Stream fluxes across \E{174}, \W{166}
and \W{158} sections may be caused by a recirculation of the Alaskan Stream waters in the
Alaska gyre.
A part of the Alaskan Stream waters inflows
into the Bering Sea through the Aleutian Passes located eastward of the Near Strait
(\E{172}~-- \W{166}).

Interannual variation in the surface flux of the Alaskan Stream across \E{174} and
the surface flux through the Near Strait correlate with changes in along Aleutian Islands wind
stress. The correlation coefficients between the zonal wind stress (\N{52}, \E{165}~-- \W{170}),
averaged for November~-- March, and annually
averaged surface flux of the Alaskan Stream (Fig.~\ref{fig4}b) and the surface flux through the Near Strait
are $-0.58$ and $-0.74$ (low-pass filtered series).
In November~-- March, the Aleutian Low
develops in the North Pacific, strong westward winds appear in the Bering Sea, and
eastward winds prevail in the northern
North Pacific (Fig.~\ref{fig3}a). An increased westward winds (negative values of the
zonal wind
stress) over the Aleutian Islands in November~-- March may force a westward
surface flow of the
Alaskan Stream and enhanced water flux via the Near Strait into the Bering Sea.

\subsection{Eddy impact on the flow through the Kamchatka Strait}

The flow pattern in the KS during summer (June~-- September) is determined by strength of
cyclonic and anticyclonic eddies in the
strait area. Passing of a cyclone through the KS enhances a
southward flux on the western side of the strait
and a northward flux on its eastern side (Fig.~\ref{fig5}d). Anticyclones
typically increase
northward flux through the western KS and southward flux through
the eastern KS. In some summers, there appears a cyclonic companion of the anticyclone in the strait.
Such a vortex pair practically blocks water exchange through the KS.
The Lagrangian map in Fig.~\ref{fig5}e demonstrates that effect in July, 2004.
There is a strong negative correlation ($r=-0.74$)
between the surface fluxes via the western and eastern parts of the strait averaged
for June~-- September (Fig.~\ref{fig6}a).

Observations of eddies in the Bering Sea basin have been made by the authors of
papers \citet{Verkhunov92,Stabeno94,Cokelet96,Mizobata02} and others. The model
results over the 26-year simulation, 1974--2000 \citep{Maslowski08} show frequent and
complex eddy activity in the Bering Sea with lifetimes of the order of a few months.
Half of the eddies
is anticyclonic and the other half is cyclonic. Diameters of those eddies are 120~km and greater
and velocities are up to 40~cm s$^{-1}$. Instabilities along the Bering Slope and the Kamchatka
Current and interactions with canyons and embayments at the landward edge of these currents, 
as well as
inflows through the Aleutian Islands Passes, may be responsible for eddy generation
in this region \citep{Kinder80,Cokelet96}.

\section{Conclusions}\label{concl}
\begin{enumerate}
\item The surface southward flux through the Kamchatka Strait demonstrates relatively high amplitude of
its seasonal variations. The southward flow 
through the strait is strong between November and April and relatively weak in June~--
September.
The strong seasonality in surface outflow through the Kamchatka Strait can be explained
by temporal changes in the wind stress
over the northern and western Bering Sea slopes.
\item The interannual changes in a surface outflow through the Kamchatka Strait statistically
significantly correlate with Near Strait inflow and Bering Strait outflow.
Enhanced westward surface flux of the Alaskan Stream across the \E{174}
section
in the northern North Pacific is accompanied by an increased inflow into the Bering Sea through
the Near Strait.
\item In summer, the surface flow pattern in the Kamchatka Strait is determined by passing of
anticyclonic and cyclonic eddies. Those eddies
are formed in the central and western Bering Sea with wind stress over the Bering basin
in winter~-- spring to be responsible for eddy generation in the region.
\end{enumerate}

\section*{Acknowledgements}
This work was supported by the Russian Foundation for Basic Research
(project nos.~11--05--98542, 12--05--00452, 13--05--00099 and 13--01--12404).
The altimeter products were distributed by AVISO with support from CNES.

\bibliographystyle{model2-names}
\bibliography{paper}   % name your BibTeX data base

\end{document}